\documentclass{article}

\usepackage{arxiv}

\usepackage{url}            % simple URL typesetting
\usepackage[ toc, page, title, titletoc ]{appendix}
\usepackage[ruled,noline]{algorithm2e}
\usepackage{amsmath}
\usepackage{amsfonts}
\usepackage[colorinlistoftodos]{todonotes}
\usepackage{hyperref}
\usepackage{lineno}
\usepackage{mathrsfs}
\usepackage{float}
\usepackage{url}            % simple URL typesetting

\usepackage{etoolbox}
\providetoggle{fullBackground}

\usepackage{parskip}  % remove indents in paragraphs

\newcommand{\DFT}{\text{DFT}}

\newcommand{\IDFT}{\text{DFT}^{-1}}

\newcommand{\FT}{\mathscr{F}}
\newcommand{\IFT}{\mathscr{F}^{-1}}

\DeclareFontFamily{U}{wncy}{}
\DeclareFontShape{U}{wncy}{m}{n}{<->wncyr10}{}
\DeclareSymbolFont{mcy}{U}{wncy}{m}{n}
\DeclareMathSymbol{\comb}{\mathord}{mcy}{"58} 
\usepackage{accents}
\newlength{\dhatheight}

\title{Least Squares Optimal Density Compensation for\\ the Gridding Non-uniform Discrete Fourier Transform}

\author{
  Nicholas Dwork\thanks{www.nicholasdwork.com, nicholas.dwork@ucsf.edu} \\
  Department of Radiology and Biomedical Imaging \\
  University of California in San Francisco
    \And
  Daniel O'Connor \\
  Department of Mathematics and Statistics \\
  University of San Francisco
    \And
  Ethan M. I. Johnson \\
  Department of Biomedical Engineering \\
  Northwestern University
    \And
  Corey A. Baron \\
  Robarts Research Institute \\
  Western University
    \And
  Jeremy W. Gordon \\
  Department of Radiology and Biomedical Imaging \\
  University of California in San Francisco
    \And
  John M. Pauly \\
  Department of Electrical Engineering \\
  Stanford University
    \And
  Peder E. Z. Larson \\
  Department of Radiology and Biomedical Imaging \\
  University of California in San Francisco
}

\begin{document}
\maketitle

\begin{abstract}
  The Gridding non-uniform Discrete Fourier Transform algorithm has shown great utility for reconstructing images from non-uniformly spaced samples in the Fourier domain in several imaging modalities.  Due to the non-uniform spacing, some correction for the variable density of the samples must be made.  Existing methods for generating density compensation values are either sub-optimal or only consider a finite set of points (a set of measure $0$) in the optimization.  This manuscript presents the first density compensation algorithm for a general trajectory that takes into account the point spread function over a set of non-zero measure.  We show that the images reconstructed with Gridding using the density compensation values of this method are of superior quality when compared to density compensation weights determined in other ways.  Results are shown with a numerical phantom and with magnetic resonance images of the abdomen and the knee.
\end{abstract}

% keywords can be removed

%\textbf{This work has been published in \textit{Magnetic Resonance Materials in Physics, Biology and Medicine}.}

\section{Introduction}
\label{sec:intro}

%\todo{Show the point spread functions}
%\todo{Use fully sampled dataset with all algorithms to create a ``silver" standard.  Then reduce the number of datapoints and see how each algorithm does.}

In several imaging modalities including Magnetic Resonance Imaging (MRI) \cite{nishimura2010principles}, Computed Tomography (CT) \cite{osullivan1985}, Spectral Domain Optical Coherence Tomography (SD-OCT) \cite{hillmann2009}, and radio astronomy \cite{pawsey1955}, a formalism exists to represent the collected data as samples in the Fourier domain.  When the samples are unique and located on Cartesian grid, e.g. spin-warp imaging with MRI, then image reconstruction can be accomplished with the inverse Discrete Fourier Transform (DFT).  However, when the samples are not located on a uniform grid, a non-uniform DFT must be used; this is the Gridding problem, a non-uniform DFT problem of type I (as defined in \cite{greengard2004accelerating}: a discrete conversion from a non-Cartesian sample set in the Fourier domain to a Cartesian sample set in the space domain).

A set of samples consists of a vector of frequencies $\boldsymbol{k}=(k_1,k_2,\ldots,k_M)\in\mathbb{R}^{M\times D}$ (called the sample coordinates) and a corresponding vector of Fourier values at those frequencies:\\ $\boldsymbol{G}=\left( G(k_1), G(k_2), \ldots, G(k_M) \right)\in\mathbb{C}^M$.  
Here, $D$ represents the dimension of the source and destination domains.
Reconstruction estimates the values of\\ $\boldsymbol{g}=\left( g(x_1), g(x_2), \ldots, g(x_N) \right)\in\mathbb{C}^N$ for a vector of space domain locations\\ $\boldsymbol{x}=(x_1,x_2,\ldots,x_N)\in\mathbb{R}^{N\times D}$.  Here, $g =\IFT G$, where the symbol $\IFT$ represents the inverse Fourier transform\footnote{The definitions of the Fourier Transform, the Discrete Fourier Transform, and their inverses are listed in Appendix \ref{sec:fourierDefs}.}.  For this paper, we make the assumption that the support of $g$ (called the Field of View) is a subset of $\boldsymbol{N}=[-N_1/2,N_1/2]\times\cdots\times[-N_D/2,N_D/2]$.  (Here, set multiplication represents the Cartesian cross product.)  If the frequencies $\boldsymbol{k}=(k_1,\ldots,k_M)$ and coordinates $\boldsymbol{x}=(x_1,\ldots,x_N)$ are equally spaced with $M=N$ (where the spacings between adjacent $k$ frequencies are inversely related to the spacings between adjacent $x$ coordinates) and both $\boldsymbol{x}$ and $\boldsymbol{f}$ include the origin, then the image can be reconstructed with the inverse DFT, which can be derived as a Riemann sum approximation to the Fourier transform \cite{epstein2008}.  When the frequencies $\boldsymbol{k}$ do not lie on a uniform grid, then the inverse DFT is no longer an appropriate reconstruction algorithm; this is the Gridding problem.

A simple (though computationally inefficient) approach to Gridding is to calculate the following sum directly \cite{epstein2008}:
\begin{equation}
  \hat{g}(x_n) = \sum_{m=0}^{M-1} w_m \, G(k_m) \exp( i\,2\pi\,k_m\cdot x_n ),
  \label{eq:slowGridding}
\end{equation}
where, $k_m\in\mathbb{R}^D$, $x_n\in\mathbb{R}^D$, and $\cdot$ represents the dot product.  The elements of the vector $w=(w_1,\ldots,w_M)\in\mathbb{R}^M$ are called the density compensation values, which can dramatically affect the quality of the reconstructed image \cite{moratal2011magnetic}.
Note that \eqref{eq:slowGridding} can be expressed as $\hat{g} = \IFT\left\{ G\,S_w\right\} = g\, \ast\, s_w$, where $\ast$ represents continuous convolution, $S_w~=~\sum_{m=0}^{M-1} w_m\,\delta(k-k_m)$, $\delta$ represents the Dirac delta function, and $s_w=\IFT\{S_w\}$.

The purpose of the values of $w$ is made clear through a simple thought experiment.
Suppose one had collected $M$ data points in the Fourier domain that were located on a Cartesian grid and should be reconstructed with the inverse DFT, which amounts to calculating \eqref{eq:slowGridding} with all $w_m$ set to $1/M$.  But then, one collects an $(M+1)^{\text{th}}$ point at the location $k_0$ a second time.  How should the estimates of $g$ be altered?  Intuitively, one would expect that the values $w_0$ and $w_{M}$ should both be $1/2$, and this would be correct.  But what should one do if, instead of collecting $k_0$, one collected $k_0+\Delta$ where $\Delta>0$ is a small deviation from the coordinate.  In this case, it is not clear what the values of $w$ should be.

%In this paper, we are concerned with the more general problem where points are collected in the Fourier domain at a finite number of unconstrained locations, and we wish to estimate the image $g$ at the locations on a Cartesian grid.

%Several methods have been proposed to calculate the density compensation values.  The most intuitive method is to calculate the Voronoi cells of each point and set the density compensation value equal to the area of the cell \cite{pawsey1955,rasche1999}.  Points that are closer together will have reduced weights while points that are farther apart will have higher weights.  Any point that is acquired multiple times has its area evenly divided to determine the weights.  But this method is suboptimal.  Consider two instances of a a radial trajectory: 1) the $0$ frequency point is acquired in each radial trajectory, while in 2) the center point is only acquired in on radial trajectory and all other radial lines are offset so that the center point lies in between two acquisitions.  In situation (1), the area of the center point will be divided by the number of radial lines in order to set the weights for the $0$ frequency points.  In situation (2), the $0$ frequency point will be given a neglibile area, for it is surrounded by other points.  This discontinuity between the weights assigned to the $0$ frequency point illustrates that t

Several methods have been proposed to determine these density compensation values.  The most intuitive method is to calculate the Voronoi cell of each point and set the density compensation values equal to the areas of the corresponding cells \cite{bracewell1973main,rasche1999,malik2005gridding} (with some exception made for the points on the convex hull).  This corresponds to estimating the inverse Fourier transform with a Riemann sum where the partition of the sum is the set of Voronoi cells.  The is an intuitive and computationally efficient approach but is not generally optimal (it does not optimize a quality metric).  And the density compensation value assigned to points on the convex hull of the sample coordinates is arbitrary.
The methods of \cite{jackson1991,pipeAbstract,pipe1999,samsonov2003determination,johnson2009convolution,zwart2012efficient} require the use of a convolution kernel, which is unnecessary given that there is no convolution kernel inherent to the summation of \eqref{eq:slowGridding}.  Moreover, these methods as well as that of \cite{song2006efficient} consider a finite set of points in the objective function (a set of measure $0$).  However, since $\hat{g}$ is a function of a continuous convolution of $g$ with $s_w$, the values of $s_w$ at all points in twice the field of view $2\boldsymbol{N}=[-N_1,N_1]\times\cdots\times[-N_D,N_D]$ are relevant.

In what follows, we present the first density compensation algorithm for a general trajectory that takes into account the point spread function over a set of non-zero measure, and we show that the images reconstructed with this method are of superior quality when compared to density compensation weights determined in other ways.

\section{Theory}
\label{sec:theory}

Note that if $s_w\approx\delta$ then $\hat{g}\approx g$.  As previously stated, we further assume that $g$ has compact support, which is a subset of $\boldsymbol{N}=[-N_1/2,N_1/2]\times[-N_2/2,N_2/2]\times\cdots\times[-N_D/2,N_D/2]$.  In this case, $s_w$ only need approximate the Dirac delta function well over $2\boldsymbol{N}=[-N_1,N_1]\times\cdots\times[-N_D,N_D]$ in order for $\hat{g}\approx g$.  This motivates our approach; we will attempt to find a density compensation vector $\boldsymbol{w}$ such that $s_w(x)\approx\delta(x)$ for all $x\in\prod_{d=1}^D [-N_d,N_d]$, where $\prod$ represents a product over a set of indices.

The function $s_w(x)=\IFT\{S_w\}=\sum_{m=1}^M w_m\,\exp(-i\,2\pi\,k_m\cdot x)$ is actually a function (as opposed to a distribution) with a finite value at all locations of the domain.  This differs from the Dirac delta function, which is actually a distribution; thus, $s_w$ could never equal the Dirac delta function over any domain including $0$.  Our hope, then, is that we can find some function that approximates the Dirac delta function well in the sense that it is a steep function centered at the origin that integrates to $1$ \cite{bracewell1995two}.

Solving optimization problem \eqref{eq:wOptProb} for some value $r>0$ finds density compensation values such that the form of the point spread function is a peak (high at the origin and low everywhere else).
\begin{equation}
  \begin{aligned}
    \underset{\boldsymbol{w}}{\text{minimize}} &\hspace{0.5em}
      \underset{2\boldsymbol{N}}{\int\int\cdots\int} 
        \exp\left(-\sum_{d=1}^D \frac{ |x_d| }{ \gamma_d } \right) \, |s_w(x)|^2 d\boldsymbol{x} \\
      \text{subject to} &\hspace{0.5em} s_w(0) = r > 0 \text{ and } w_m \ge 0 \text{ for all } m.
  \end{aligned}
  \label{eq:wOptProb}
\end{equation}
One constraint of \eqref{eq:wOptProb} sets the value of the point spread function at $x=0$ to be $r>0$; together with the objective function the problem attempts to find values that yield a peak near the origin and small values away from the origin.  The values are constrained to be non-negative to make the interpolation process of \eqref{eq:slowGridding} more stable.
The $\exp\left(-\sum_{d=1}^D |\cdot| / \gamma_d \right)$ function is a weighting that makes \eqref{eq:wOptProb} a constrained weighted least squares problem.  Because the value at the origin is constrained to $r$, and because the resulting $s_w$ function is continuous, it will be more difficult to make values close to the origin small than it is to make values distant from the origin small.  This weighting permits the user (through the use of the parameter $\gamma>0$) to trade off between some additional error in areas more distant from the origin with significant improvements in those values close to the origin.

Consider the constraint $s_w(0)=r$.  Note that $s_w(0) = \boldsymbol{1}^T w = r$, an equivalent simpler expression.
We are left to determine the value of $r$.  Moreover, solving problem \eqref{eq:wOptProb} does not yield a point spread function that integrates to $1$, which is the remaining requirement of a good approximation of a Dirac delta function.  Let us consider two different values of $r$ for \eqref{eq:wOptProb}: $1$ and $r'$.
Let $w'$ be the solution to problem \eqref{eq:wOptProb} when $r=r'$.  Let $\tilde{w}=w' / r'$.  Since each component of $w'\ge 0$, every component of $\tilde{w}\ge 0$.  Additionally, $\boldsymbol{1}^T \tilde{w} = \boldsymbol{1}^T w' / r' = 1$.  Finally, let us determine the value of the objective function when $w=\tilde{w}$.  In this case, $|s_{\tilde{w}}(x)|^2 = (r')^2 |s_{w'}(x)|^2$.  Thus, the solution when $r=1$ is simply $\tilde{w}=w'/r'$.

More generally, problem \eqref{eq:wOptProb} can be solved with $r=1$ and then scaled to attain the solution for other values of $r$.  So there is a unique value of $\kappa = 1/r'$ such that the integral over $\boldsymbol{\eta}$ is $1$, where $\boldsymbol{\eta}$ is a small rectangular region centered at the origin:
\begin{equation}
  \underset{\boldsymbol{\eta}}{\int\int\cdots\int} s_{\kappa w}(x) \, d\boldsymbol{x} = 1.
  \label{eq:diracConstraint}
\end{equation}
Thus, to find the density compensation values, one solves problem \eqref{eq:wOptProb} with $r=1$ and then scales the weights by $\kappa=1/r'$ such that \eqref{eq:diracConstraint} is satisfied.  The expression for $r'$, derived in Appendix \ref{sec:rPrime}, is
\begin{equation*}
  r' = \left( \sum_{j=1}^M w_j \Pi_{d=1}^D \left(\frac{\sin(\pi \, k_{j,d} \, \eta_d)}{\pi \, k_{j,d}}\right) \right)^{-1}.
\end{equation*}

This set of scaled density compensation values are optimal in the sense that they solve the following optimization problem:
\begin{equation}
  \begin{aligned}
    \underset{w}{\text{minimize}} &\hspace{0.5em}
      \underset{2\boldsymbol{N}}{\int\int\cdots\int} 
        \exp\left(-\frac{1}{\gamma}\sum_{d=1}^D |x_d|\right) \, |s_w(x)|^2 d\boldsymbol{x} \\
      \text{subject to} &\hspace{0.5em} s_w(0) = r' \text{, } w_m \ge 0 \text{ for all } m, \\
      \text{and} &\hspace{0.5em} \underset{\boldsymbol{\eta}}{\int\int\cdots\int} s_{w}(x) \, d\boldsymbol{x} = 1.
  \end{aligned}
\end{equation}

\section{Algorithm}
\label{sec:alg}

To solve problem \eqref{eq:wOptProb}, we convert it into a form that can be solved with a gradient-projection algorithm.  Let $f_0$ denote the objective function of the problem.  And let $\mathbb{I}_{Pr}$ denote the indicator function that the argument is an element of the probability simplex (equal to $0$ when the argument is an element of the probability simplex and equal to infinity otherwise).  Then problem \eqref{eq:wOptProb} is equivalent to
\begin{equation*}
  \underset{w}{\text{minimize}} \hspace{0.5em} f_0(w) + \mathbb{I}_{Pr}(w).
\end{equation*}

To solve this problem with the gradient-projection algorithm, we need an implementation of the gradient of $f_0$, denoted $\nabla f_0$ and an implementation of the Euclidean projection onto the probability simplex, denoted $\Pi_{Pr}$ (which is the proximal operator of the corresponding indicator function \cite{parikh2014proximal}).  We use the efficient algorithm of Wang and Carreira-Perpin{\'a}n for this projection operator \cite{wang2013projection}.

The gradient of the objective function, derived in appendix \ref{sec:objectiveGradient}, is $\nabla f_0(w) = A w$ where
\begin{equation*}
  A_{lj} = \left\{ \begin{array}{ll}
    2\prod_{d=1}^D 2\,N_d & \text{ if } k_{j,d} = k_{l,d} \\ \\
    2\prod_{d=1}^D \frac{1}{\pi(k_{j,d}-k_{l,d})}\sin\left( 2\pi \left(k_{j,d}-k_{l,d}\right) N_d \right)  & \text{ otherwise }
    \end{array}  \right. .
\end{equation*}

With the above definitions, problem \eqref{eq:wOptProb} can be solved by alternating between an iteration of gradient descent and a Euclidean projection onto the probability simplex, as shown in Alg. \ref{alg:gradProj}.  The values of $x^{(0)}$ are initialized to a constant vector with values of $1/M$.  For faster convergence, we used the Fast Iterative Shrinkage Threshold Algorithm (FISTA) \cite{beck2009fast}.  To take advantage of regions where the local gradient is much smaller than its global bound, we used FISTA with line search \cite{scheinberg2014fast}.  The norm of $A$ is estimated using power iteration, and the initial step size of the optimization algorithm is set to $1/\|A\|$.

\begin{algorithm}[ht]
    \protect\caption{Gradient-Projection Algorithm}
    \label{alg:gradProj}

    \textbf{Inputs:}  $K > 0$, $\mu>0$, \text{ and } $x^{(0)}$.

    \textbf{For} $k = 1, 2, \ldots, K$

    \hspace{1em}  $y^{(k)} = x^{(k-1)} - \mu \, \nabla f_0\left( \, x^{(k-1)} \right)$

    \hspace{1em}  $x^{(k)} = \Pi_{Pr}\left(\, y^{(k)} \,\right)$

    \textbf{End For}

    \textbf{Outputs: } $x^{(K)}$
\end{algorithm}

For reference, we call the results of this algorithm the \textit{Gradient Projection} density compensation values.

\section{Experiments}
\label{sec:experiments}

\begin{figure}[ht]
  \centering{}
  \includegraphics[width=0.8\linewidth]{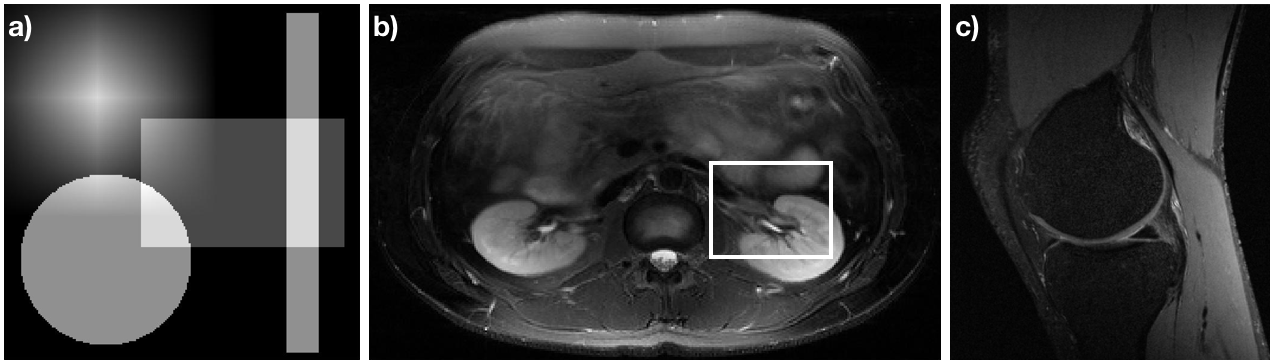}
  \caption{Accurate images for data analyzed in this manuscript.  (a) A numerical phantom consisting a the sum of a separable tri function, a circ function, and two separable stretched rect functions, each translated from the origin.  (b) Slice 6 of the Double Vision dataset from the 2010 reconstruction challenge of the ISMRM.  (c) A slice of a knee acquired from \url{mridata.org}.  }
  \label{fig:trueImgs}
\end{figure}

We present results for three experiments with two-dimensional data: (1) a numerical phantom, and (2) multi-coil magnetic resonance (MR) data from the 2010 reconstruction challenge of the International Society of Magnetic Resonance in Medicine (ISMRM) \cite{pipeReconChallenge}, and (3) MR data of a slice of a knee.  Fourier domain data was normalized to lie in the $[-0.5,0.5]^2$ square.  The true images of the experiments can be seen in Fig. \ref{fig:trueImgs}.  Once the density compensation values were determined, rather than calculating the sum of \eqref{eq:slowGridding} directly, we use the more efficient algorithm of \cite{osullivan1985,jackson1991,beatty2005} with a Kaiser-Bessel kernel and an oversampling ratio of $1.5$.  For all results of the gradient projection algorithm, the value of $\gamma$ used was $25\%$ of the length the side of the image; this value was chosen to ensure that there was significant weighting near the center of the point spread function and to ensure that the weightings near the boundary of $2\boldsymbol{N}$ were non-negligible.  The sizes of the images of the phantom, the abdomen, and the knee were, $208\times 208$, $190\times 305$, and $300\times 270$, respectively.  Correspondingly, the values of $(\gamma_1,\gamma_2)$ were $(52,52)$, $(47.5,76.25)$, and $(75,67.5)$, respectively.

\begin{figure}[ht]
  \centering{}
  \includegraphics[width=0.7\linewidth]{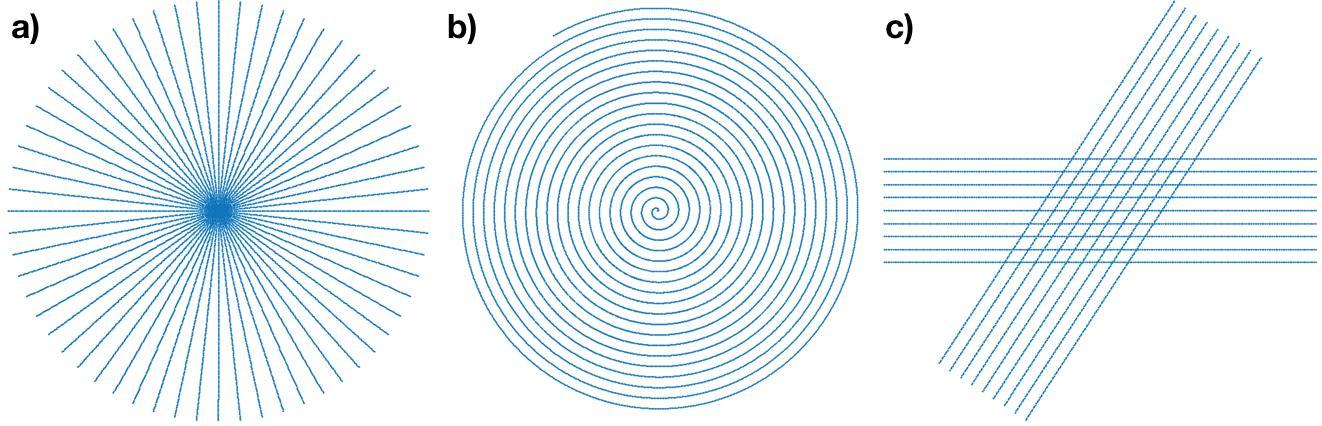}
  \caption{Subsets of the sample coordinates used for the data of Fig. \ref{fig:trueImgs}.  Sub-image (a) shows every $6^{\text{th}}$ radial line for the radial sample set used with the numerical phantom Fig. \ref{fig:trueImgs}a.  Sub-image (b) shows $1$ of the eight spiral interleaves from the sample set used with Fig. \ref{fig:trueImgs}b.  Sub-image (c) shows $2$ of the $60$ angles from the propeller sample set used with Fig. \ref{fig:trueImgs}c. }
  \label{fig:trajectories}
\end{figure}

The numerical phantom, shown in Fig. \ref{fig:trueImgs}a, consists of a separable tri function,  a circ function, and two separable scaled rect functions all offset from the origin and summed together.  The Fourier values of this phantom are known.  (The Fourier transform of a tri function is $\text{sinc}^2$, the Fourier transform of a circ is a jinc, and the Fourier transform of a rect is a sinc.  When combined with the Fourier shift and scaling theorems, the Fourier values of the numerical phantom can be determined analytically.)  The sample coordinates used with this phantom are radial with $360$ spokes, $150$ points per spoke.  An image of every sixth spoke from this sample set is shown in Fig. \ref{fig:trajectories}a.

The image shown in Fig. \ref{fig:trueImgs}b consists of an axial slice of an abdomen with an $8$ coil acquisition from the 2010 reconstruction challenge of the International Society of Magnetic Resonance in Medicine \cite{pipeReconChallenge}.  (Slice $6$ of the $12$ slice acquisition is used in this manuscript.)  The method of Roemer et al. is used to combine the individual images of each coil into a single image for display \cite{roemer1990nmr}.  The spiral trajectory created by Craig Meyer for the challenge was used, which consists of $8$ spiral interleaves with $19$ revolutions per interleave.  However, only every $5^\text{th}$ sample of the readout was retained for a total of $4000$ points per interleave.
A single interleave of this set of sample coordinates is shown in Fig. \ref{fig:trajectories}b.

The image shown in Fig. \ref{fig:trueImgs}c consists of a sagittal slice of a knee acquired from \url{mridata.org}.
The coordinates of the sample set used for analysis is the propeller trajectory \cite{pipe1999motion} with $90$ angles of acquisition (with $2$ degrees difference between adjacent angular acquisitions), $9$ lines per angle separated by $0.03$, $200$ points per line, and $60$ angles.  Two angles for the sample set are shown in Fig. \ref{fig:trajectories}c.

For the images of the abdomen and the knee, a type II non-uniform DFT was used to estimate the Fourier values given the images \cite{pruessmann2001advances}.  Note that inverse Gridding is not the inverse operation of Gridding, which is not generally invertible.  In the case of the abdomen, inverse gridding was applied to the image from each coil separately.

Gridding reconstructions were done with density compensation values determined using the Voronoi cell based algorithm of \cite{rasche1999}, the fixed point (FP) algorithm of \cite{pipeAbstract,pipe1999} with a Kaiser-Bessel kernel, and the proposed gradient projection (GP) algorithm.  For FP, the kernel was normalized so that it integrates to $1$ in order to scale the reconstructed image correctly.  The mean square error (MSE) metric and the structural similarity metric (SSIM) was calculated for the difference between the reconstruction and the true image.

All code was written in Matlab R2019b and computations were performed by a $4$-core 2010 Macbook Pro.  The multiplications of the matrix $A$ and its transpose were parallelized across the $4$ cores.

\section{Results}
\label{sec:results}

\begin{figure}[ht]
  \centering{}
  \includegraphics[width=0.8\linewidth]{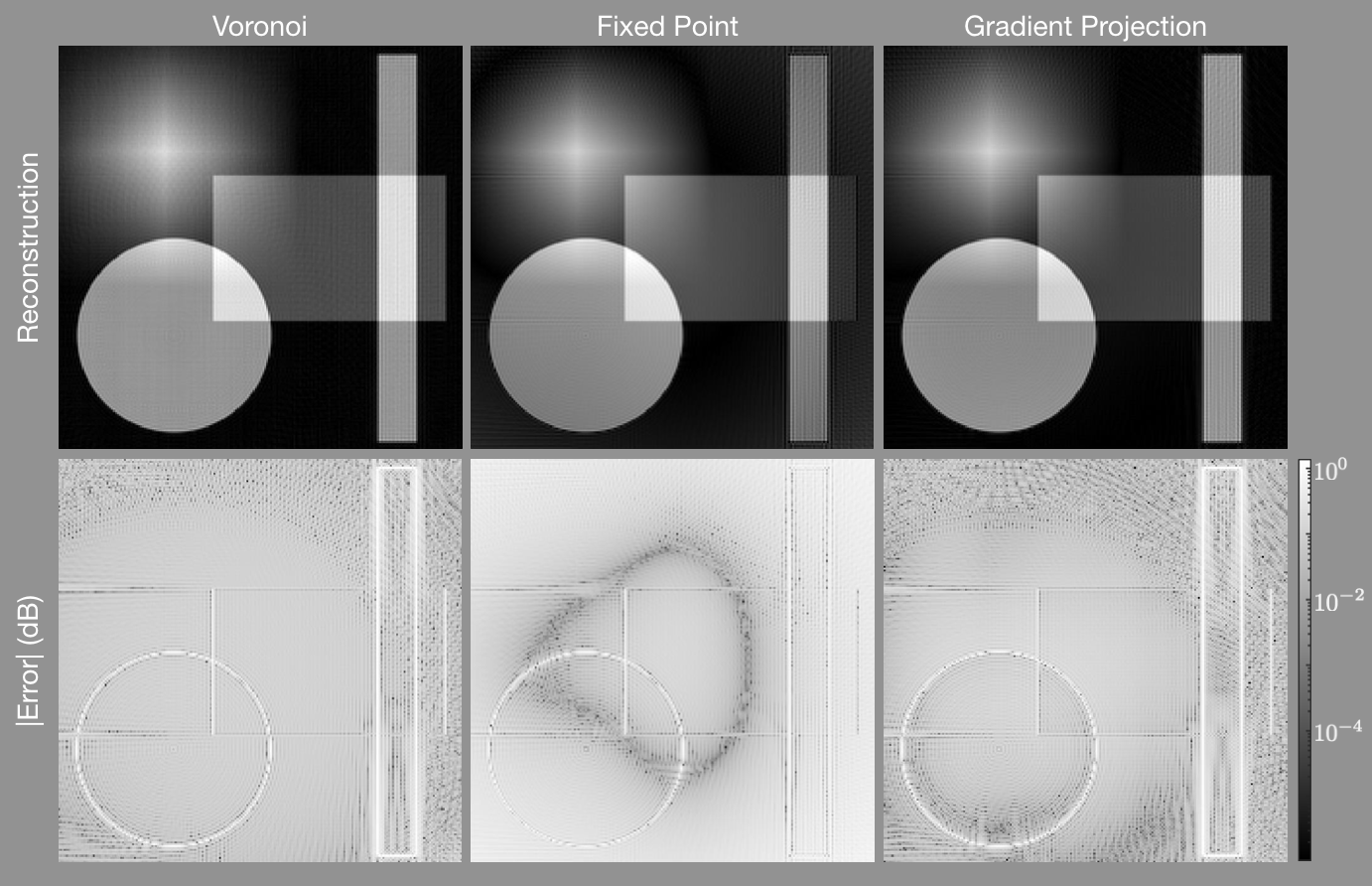}
  \caption{Gridding reconstructions of the phantom with density compensation values determine using (left) the areas of Voronoi cells, (center) the fixed-point iteration algorithm, and (right) the gradient projection algorithm.  The first row shows the reconstructed images, the second row shows the error of each pixel in decibels.  The gradient projection algorithm has lower error in more portions of the image than the other techniques. }
  \label{fig:phantomResults}
\end{figure}

\begin{figure}[ht]
  \centering{}
  \includegraphics[width=0.8\linewidth]{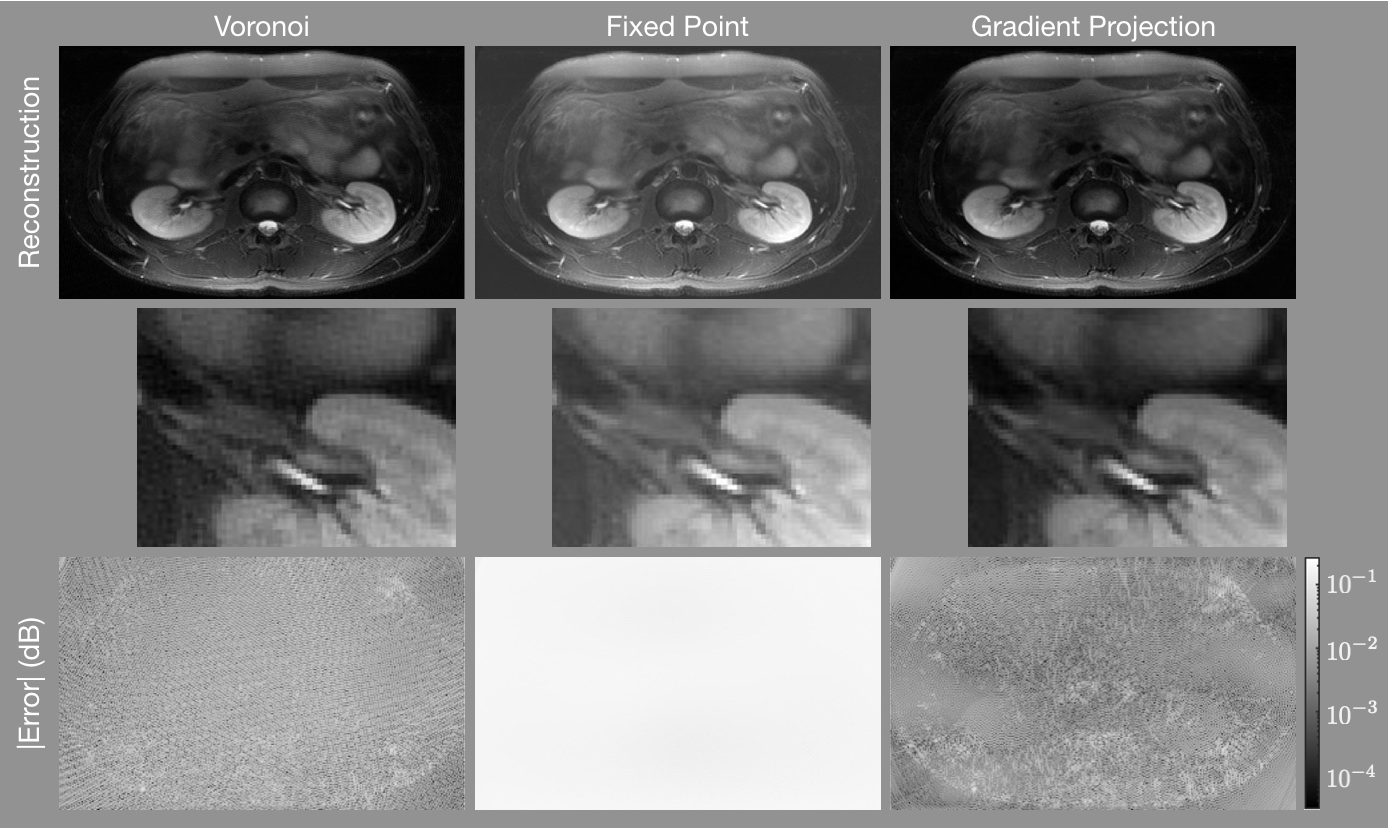}
  \caption{Gridding reconstructions of the abdomen with density compensation values determine using (left) the areas of Voronoi cells, (center) the fixed-point iteration algorithm, and (right) the gradient projection algorithm.  The first row shows the reconstructed images, the second row shows a zoomed into a portion of the image centered on the left kidney indicated by the white rectangle shown in Fig. \ref{fig:trueImgs}b, and the third row shows the error of each pixel in decibels.  The gradient projection algorithm has lower error in more portions of the image than the other techniques.}
  \label{fig:abdomenResults}
\end{figure}

\begin{figure}[ht]
  \centering{}
  \includegraphics[width=0.8\linewidth]{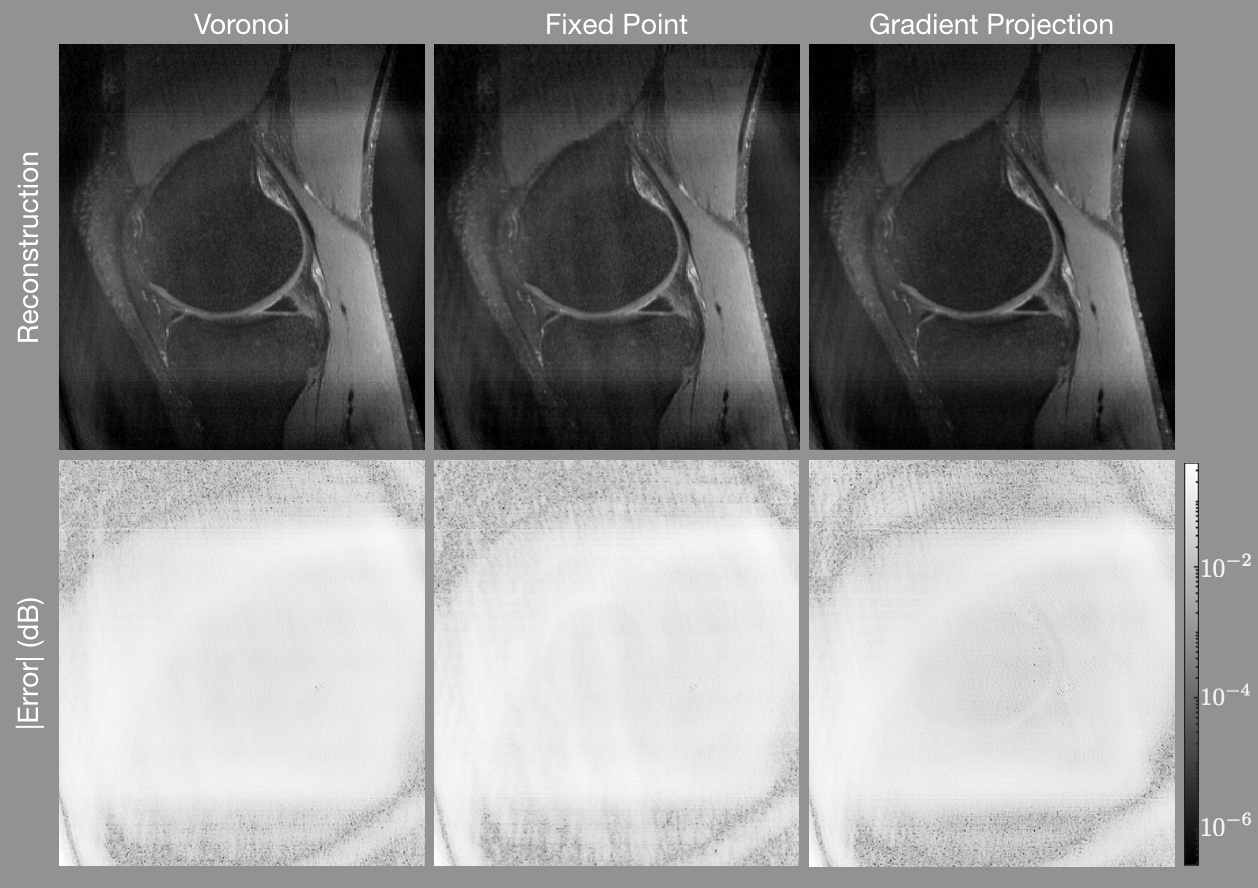}
  \caption{Gridding reconstructions of the knee with density compensation values determine using (left) the areas of Voronoi cells, (center) the fixed-point iteration algorithm, and (right) the gradient projection algorithm.  The first row shows the reconstructed images and the second row shows the error of each pixel in decibels.  The gradient projection algorithm has lower error in more portions of the image than the other techniques.}
  \label{fig:kneeResults}
\end{figure}

Figures \ref{fig:phantomResults}, \ref{fig:abdomenResults}, and \ref{fig:kneeResults} show magnitude reconstructions for the phantom, the abdomen, and the knee, respectively.  In Fig. \ref{fig:phantomResults}, the top row shows the reconstructions of each algorithm and the bottom row shows the error image in decibels.   All reconstruction algorithms generated the most significant errors at the edges of the circ and rect functions.  Overall, the GP algorithm has less error than the other methods, as evidenced by the dark regions in the difference image.

Figure \ref{fig:abdomenResults} shows results for the abdomen.  The top row shows the reconstructed images, the middle row shows a region centered near the left kidney zoomed into the white rectangle shown in Fig. \ref{fig:trueImgs}b, and the bottom row shows the difference images.
As with the numerical phantom, there are regions of the image with less error when using the gradient projection algorithm than when using the competing algorithms.  The FP reconstruction yields an image that is too bright.  Additionally, it alters the contrast of the image significantly, making regions of the center of the image much brighter than the true image.  In the zoomed-in images, note that there is an erroneous high frequency block pattern super-imposed on the reconstruction with the Voronoi cell based weights that is absent in the GP reconstruction.

Table \ref{tbl:qualityMetrics} shows the mean square error, the structural similarity metric \cite{wang2004image}, and the runtimes for for generating the density compensation values.  In all cases, the gradient projection algorithm yields the lowest (best) mean square error and the highest (best) structural similarity value.  The Voronoi method is the fastest method, followed by FP.  The GP method takes dramatically longer than either of the other methods.

\begin{table}[h!]
  \caption{Mean square error and structural similarity metric of each reconstruction}
  \begin{center}
  \begin{tabular}{ |c|c|c|c| } 
    \hline
    Image & Voronoi & Fixed Point & Gradient Projection \\
    \hline
    & \multicolumn{3}{|c|}{Mean Square Error} \\
    \hline
    Phantom & $0.0282$ & $0.075$ & $0.0237$ \\
    Abdomen & $3.2(10^{-4})$ & $0.031$ & $1.0(10^{-4})$ \\
    Knee & $1.1(10^{-2})$ & $1.2(10^{-2})$ & $0.72(10^{-2})$ \\
    \hline
    & \multicolumn{3}{|c|}{Structural Similarity Metric} \\
    \hline
    Phantom & $0.986$ & $0.954$ & $0.986$ \\
    Abdomen & $0.995$ & $0.800$ & $0.998$ \\
    Knee & $0.807$ & $0.798$ & $0.846$ \\
    \hline
    & \multicolumn{3}{|c|}{Runtime (seconds)} \\
    \hline
    Phantom & 1.1 & 19.2 & 3097 \\
    Abdomen & 38 & 238 & 20607 \\
    Knee & 3.1 & 149 & 15312 \\
    \hline
  \end{tabular}
  \end{center}
  \label{tbl:qualityMetrics}
\end{table}

%Figure \ref{fig:kneeGammas} shows how the reconstruction of the knee with density compensation values determined using GP changes as the value of $\gamma$ changes.  Table \ref{tbl:kneeGammas} shows how the mean square error and the structural similarity metric change with as the value of $\gamma$ changes.  

%\begin{figure}[ht]
%  \centering{}
%  \includegraphics[width=0.8\linewidth]{./kneeGammas.png}
%  \caption{Knee gammas}
%  \label{fig:kneeGammas}
%\end{figure}

%\begin{table}[h!]
%  \caption{Mean square error and structural similarity metric of GP reconstructions of the knee}
%  \begin{center}
%  \begin{tabular}{ |c|c|c|c| } 
%    \hline
%    $\gamma$ & 0.1 & 0.25 & 0.5 \\
%    MSE & 9.3(10^{-2}) & 7.2(10^{-2}) & 9.6(10^{-2}) \\
%    SSIM & 0.822 & 0.846 & 0.817 \\
%    \hline
%  \end{tabular}
%  \end{center}
%  \label{tbl:kneeGammas}
%\end{table}

\section{Discussion}
\label{sec:discussion}

The results show that the quality of the reconstruction is improved when the gradient projection method is used to determine the density compensation values, as opposed to the Voronoi method or the fixed point method.  As previously stated, the density compensation values of the points that lie on the convex hull of the sample coordinates is arbitrary with Voronoi.  This may lead to the high frequency noise observed in Fig. \ref{fig:abdomenResults}.  The fixed point method is not, generally, guaranteed to converge to a set of values.  Since there is no constraint, the values may increase in size due to instabilities.  Early stopping of $8$ iterations is employed to prevent this from happening.  However, this would yield sub-optimal values.  This, along with the scaling ambiguity of the convolution kernel, may explain the significantly higher error of the fixed point iteration observed in Fig. \ref{fig:abdomenResults}b.

The gradient projection algorithm proposed is much more computationally intensive than either of the others.  Despite this long runtime, the gradient projection method is useful for systems where the sample coordinates are known prior to imaging (e.g. MRI or CT), or where the time required to collect the data is so long that the computation time is negligible in comparison (e.g. radio astronomy).  If the sample coordinates are known, the density compensation values can be computed prior to imaging and stored for future use.  In this case, the runtime does not alter the time between data acquisition and display.  If the acquisition time is so long that the computation time is negligible, in comparison, then perhaps the gradient projection method is appropriate to get the highest quality image possible from a Gridding reconstruction.

Aside from reconstructing the image directly for viewing, the result of Gridding can be used to initialize an iterative model based reconstruction algorithm \cite{fessler2010model,noel2018evaluation} such as compressed sensing \cite{candes2008introduction,lustig2007sparse,dwork2021utilizing}.  This is especially important for non-convex model based reconstruction algorithms, such as ENLIVE \cite{holme2019enlive} or MCCS \cite{dwork2020calibrationless}, since a closer initial guess reduces the probability of yielding a final answer from an erroneous local minima.

%Though the quality of the result is dependent on the value of $\gamma$, a setting of $\gamma=0.25$ generates good quality for the images studied.  Though a more thorough investigation of this parameter (on a large set of data) is necessary to draw a meaningful conclusion, the slowly varying weight function indicates that this value is likely to generate good results for most datasets. 

\section{Conclusions}
\label{sec:conclusions}

In this work, we present a method to determine the density compensation values of the Gridding non-uniform DFT.
The method for determining the density compensation values results from an optimization problem that considers twice the field of view of the object to be imaged.  This is the first method for determining the density compensation values that is optimized over a set of non-zero measure.

Though results are shown for two-dimensional data, the straightforward modification to three-dimensional data may be useful for non-Cartesian three-dimensional MRI datasets where the trajectories do not lie in a single pain, such as cones \cite{gurney2006design,wu2013free} or yarn ball \cite{stobbe2021three}.

Though this approach has been demonstrated on non-uniform discrete Fourier transform problems of type I, it may be more general.  The approach may be valid for problems of type III, where the points of both the source and the destination domain do not lie on a uniform grid.  This possibility comes from the fact that a discrete set of points in the destination domain was never used when determining the density compensation values.  The relevant change may simply be to alter the set $\{x_n : n = 1, 2, \ldots, N\}$ where the summation of \eqref{eq:slowGridding} is calculated.  We leave this pursuit as future work.

\section*{Acknowledgments}
ND has received post-doctoral training funding from the American Heart Association (grant number 20POST35200152).
ND has received funding from the Quantitative Biosciences Institute at UCSF (no grant number).
JG has received funding from the National Institute of Health / National Institute of Biomedical Imaging and Bioengineering (grant number U01EB026412).
PL has received funding from the the National Institute of Health (grant number NIH R01 HL136965).

No conflicts of interest, financial or otherwise, are declared by the authors.

\appendix

\section{Fourier Transform Definitions}
\label{sec:fourierDefs}

For this document, the Fourier Transform and Inverse Fourier Transform are defined as
\begin{equation*}
  \begin{aligned}
    F(k) &= \FT \{f\}(k) = \int_{-\infty}^{\infty} f(x) \exp\left( -i\,2\pi\,k\,x \right) dx, \\
    f(x) &= \IFT \{F\}(x) = \int_{-\infty}^{\infty} F(k) \exp\left( i\,2\pi\,k\,x \right) dk.
  \end{aligned}
\end{equation*}
The Discrete Fourier Transform and Inverse Discrete Fourier Transform are defined as
\begin{equation*}
  \begin{aligned}
    V[m] &= \DFT(v)[m] = \sum_{n=0}^{N-1} v[n] \exp\left( -i\,2\pi\,\frac{m\,n}{N}\right), \\
    v[n] &= \IDFT(V)[n] = \frac{1}{N}\sum_{m=0}^{M-1} V[n] \exp\left( i\,2\pi\,\frac{m\,n}{N}\right).
  \end{aligned}
\end{equation*}

\section{Expression for $r'$}
\label{sec:rPrime}

Here we derive the expression for the scalar multiple that satisfies constraint \eqref{eq:diracConstraint} as explained in section \ref{sec:theory}.
Recall that $\tilde{w}$ is the solution to problem \eqref{eq:wOptProb} when $r=1$.  Starting from constraint \eqref{eq:diracConstraint},
\begin{equation*}
  \begin{aligned}
    1 &= \underset{\boldsymbol{\eta}}{\int\int\cdots\int} s_{w^\star}(x) \,d\boldsymbol{x}
      = \underset{\boldsymbol{\eta}}{\int\int\cdots\int} r' \, s_{\tilde{w}}(x) \,d\boldsymbol{x} \\
      &= r' \underset{\boldsymbol{\eta}}{\int\int\cdots\int}
        \sum_{j=1}^M \tilde{w}_j \exp \left( -i\,2\pi k_j\cdot x \right) \,d\boldsymbol{x} \\
      &= r' \sum_{j=1}^M \tilde{w}_j \prod_{d=1}^D \int_{-\eta_d/2}^{\eta_d/2} \exp(-i\,2\pi k_{j,d} x_d ) \,dx_d. \\
    \text{Therefore, } r' &= \left( \sum_{j=1}^M \tilde{w}_j \, \prod_{d=1}^D \left(\frac{\sin(\pi k_{j,d} \eta_d)}{\pi k_{j,d}}\right) \right)^{-1}.
  \end{aligned}
\end{equation*}

\section{Gradient of the Objective function}
\label{sec:objectiveGradient}

Consider the objective function $f_0(w)= \int\cdots\int_{2\boldsymbol{N}} h(x,w) d\boldsymbol{x}$, where
\begin{equation*}
  h(x,w) = \exp\left(-\sum_{d=1}^D \frac{|x_d|}{\gamma_d}\right) \, |s_w(x)|^2
         = \exp\left(-\sum_{d=1}^D \frac{|x_d|}{\gamma_d}\right) \, s_w(x) \overline{ s_w(x) }.
\end{equation*}
Here, $\overline{\cdot}$ denotes the complex conjugate.
Then $\nabla f_0(w) = \int\cdots\int_{2\boldsymbol{N}} \nabla_w h(x,w) \, d\boldsymbol{x}$.
By the product rule of differentiation,
\begin{equation*}
  \begin{aligned}
    \frac{\partial}{\partial w_l} &h(x,w) = \exp\left(-\sum_{d=1}^D \frac{|x_d|}{\gamma_d}\right) \left[
        \left( \frac{\partial}{\partial w_l} s_w(x) \right)\left( \overline{ s_w(x) } \right) +
        \left( s_w(x) \right)\left( \frac{\partial}{\partial w_l} \overline{ s_w(x) } \right)
        \right] \\
      & = \exp\left(-\sum_{d=1}^D \frac{|x_d|}{\gamma_d}\right) \left[
        \sum_{j=1}^M w_j \, \exp\left( i 2\pi (k_j - k_l)\cdot x \right) +
        \sum_{j=1}^M w_j \, \exp\left( -i 2\pi (k_j - k_l)\cdot x \right)
        \right] \\
      & = 2 \exp\left(-\sum_{d=1}^D \frac{|x_d|}{\gamma_d}\right) \sum_{j=1}^M \cos\left( 2\pi (k_j - k_l)\cdot x \right).
  \end{aligned}
\end{equation*}
With this expression, we can now find an analogous expression for the partial derivative of the objective function:
\begin{equation*}
  \frac{\partial}{\partial w_l} f_0(w) = 2\sum_{j=1}^M w_j\,
    \underset{2\boldsymbol{N}}{\int\int\cdots\int} \,\cos\left( 2\pi (k_j - k_l )\cdot x \right) 
      \exp\left(-\sum_{d=1}^D \frac{|x_d|}{\gamma_d}\right) d\boldsymbol{x} = a_l \cdot w,
\end{equation*}
where $a_l$ is the vector such that the $j^{\text{th}}$ component is 
\begin{equation*}
  a_{lj} = 2\int\int\cdots\int_{2\boldsymbol{N}}\cos\left( 2\pi (k_j - k_l )\cdot x \right)
    \exp\left(-\sum_{d=1}^D |x_d|/\gamma_d \right) d\boldsymbol{x}.
\end{equation*}
We are left to determine $a_l$:
\begin{equation*}
  \begin{aligned}
    a_{lj} &= \exp\left(-\sum_{d=1}^D \frac{|x_d|}{\gamma_d}\right) \underset{2\boldsymbol{N}}{\int\int\cdots\int} \exp\left( i\,2\pi\,(k_j-k_l)\cdot x \right) d\boldsymbol{x} \\
      & \hspace{1em} + \exp\left(-\sum_{d=1}^D \frac{|x_d|}{\gamma_d}\right) \underset{2\boldsymbol{N}}{\int\int\cdots\int} \exp\left( -i\,2\pi\,(k_j-k_l)\cdot x \right) d\boldsymbol{x} \\
      &= \prod_{d=1}^D \int_{-N_d}^{N_d} \exp\left( i\,2\pi (k_{j,d}-k_{l,d})x_d - |x_d|/\gamma_d \right) \, dx_d \\
      & \hspace{1em} + \prod_{d=1}^D \int_{-N_d}^{N_d} \exp\left( -i\,2\pi (k_{j,d}-k_{l,d})x_d - |x_d|/\gamma_d \right) \, dx_d \\
      &= 2\,\text{Real}\left\{ \prod_{d=1}^D \int_{-N_d}^{N_d} \exp\left( i\,2\pi (k_{j,d}-k_{l,d})x_d - |x_d|/\gamma_d \right) dx_d \right\}.
  \end{aligned}
\end{equation*}
Consider $t_d = \int_{-N_d}^{N_d} \exp\left( i\,2\pi (k_{j,d}-k_{l,d})x_d - |x_d|/\gamma_d \right) dx_d$.  If $k_{j,d} = k_{l,d}$, then \\ $t_d=2\gamma_d\left( 1 - \exp(1 - N_d/\gamma_d) \right)$.  Otherwise,
\begin{equation*}
  \begin{aligned}
    t_d &= \int_0^{N_d} \exp\left( i\,2\pi\,(k_{l,d}-k_{j,d})\,x_d - x_d/\gamma_d \right) + 
           \int_{-N_d}^0 \exp\left( i\,2\pi\,(k_{l,d}-k_{j,d})\,x_d + x_d/\gamma_d \right) \\
        &= \frac{1}{i\nu_d - 1/\gamma_d}\exp(i\nu_d N_d - N_d/\gamma_d) - \frac{1}{i\nu_d -1/\gamma_d} \\
        &\hspace{1em} + \frac{1}{i\nu_d + 1/\gamma_d} - \frac{1}{i\nu_d + 1/\gamma_d}\exp(-i\nu_d N_d - N_d\gamma_d) \\
        &= 2\,\text{Real}\left\{ \frac{-1/\gamma_d - i\nu_d}{(1/\gamma_d)^2 + \nu_d^2} \exp\left( i\nu_d N_d -N_d/\gamma_d \right) +
           \frac{1/\gamma_d -i\nu_d}{(1\gamma_d)^2 + \nu_d^2}\right\} \\
        &= \frac{2}{(1+\gamma_d^2\nu_d^2)} \left[ \exp(-N_d/\gamma_d)\left(-\gamma_d\cos(\nu_d\,N_d) +
           \gamma^2\nu_d\sin(\nu_d\,N_d) \right) + \gamma_d \right],
  \end{aligned}
\end{equation*}
where $\nu_d=2\pi\left( k_{l,d}-k_{j,d} \right)$.

Therefore, $a_{l,j}=2\prod_{d=1}^D t_d$ where
\begin{equation*}
  t_d = \left\{ \begin{array}{ll}
    2\gamma \left( 1 - \exp(-N_d/\gamma_d) \right) & \text{ if } k_{j,d} = k_{l,d} \\ \\
    \frac{2}{1+\gamma_d^2\nu_d^2} \left[ \exp(-N_d/\gamma_d)\left( -\gamma_d\cos( \nu_d N_d ) +
      \gamma^2 \nu_d \sin(\nu N_d) \right) + \gamma \right] & \text{ otherwise }
    \end{array}  \right. .
\end{equation*}
With this expression of $a_l$, $(\partial/\partial w_l) f_0(w) = a_l \cdot w$.  Let $A$ be the matrix such that the $l^{\text{th}}$ row of $A$ is $a_l$.  Then $\nabla f_0(w) = A w$.

%\bibliographystyle{unsrt} %Used BibTeX style is unsrt
%\bibliography{references}

\end{document}